\documentclass[12pt]{article}

\usepackage{epsf,cite,a4wide,latexsym}
\newcommand{\beq}{\begin{equation}}
\newcommand{\eeq}{\end{equation}}
\newcommand{\bea}{\begin{eqnarray}}
\newcommand{\eea}{\end{eqnarray}}
\newcommand{\bda}{\begin{eqnarray*}}
\newcommand{\eda}{\end{eqnarray*}}

\begin{document}

\vskip -4cm

\begin{flushright}
MIT-CTP-3110
\end{flushright}

\vskip 0.2cm

{\Large
\centerline{{\bf SU(2) vortex configuration in laplacian center gauge}}
\vskip 0.3cm

\centerline{ \'Alvaro Montero }}
\vskip 0.3cm

\centerline{Center for Theoretical Physics}
\vskip -0.15cm
\centerline{Laboratory for Nuclear Science and Department of Physics}
\vskip -0.15cm
\centerline{Massachusetts Institute of Technology}
\vskip -0.15cm
\centerline{Cambridge, Massachusetts 02139}
\vskip -0.15cm
\centerline{USA}
\vskip 0.1cm
\centerline{e-mail: montero@lns.mit.edu}
\vskip 0.8cm

\begin{center}
{\bf ABSTRACT} 
\end{center}
We study how Laplacian Center Gauge identifies the vortex content of a thick
SU(2) vortex configuration on the lattice. This configuration is a solution of
the Yang-Mills classical equations of motion having vortex properties. We find
that this gauge fixing procedure cleanly identifies the underlying vortex properties.
We also study the monopole content of this configuration detected with this procedure.
We obtain two monopole curves lying on the surface of the vortex.


\newpage

\section{Introduction}

Confinement of quarks is still a phenomena not fully understood. Two mechanisms, proposed
long time ago, are currently receiving a lot of attention. In the first one \cite{monopoles},
confinement is seen as a dual Meissner effect, based in the condensation of magnetic 
monopoles in the QCD vacuum. In the second one \cite{vortices}, confinement is due to 
the condensation of vortices. Both pictures of confinement show up in specific partial 
gauge fixings.

In the dual superconductor picture of confinement, magnetic monopoles appear 
as defects in the abelian gauges proposed by 't Hooft \cite{abproj}. In this case
the gauge is fixed up to the Cartan subgroup of the gauge group. Then, monopoles
appear at points in space in which the gauge can not be fixed up to the Cartan
subgroup, leaving a gauge freedom larger than the abelian subgroup.
In the vortex picture of confinement vortices are bi-dimensional objects carrying
flux quantized in elements of the center of the group.

Both pictures of confinement receive strong support from lattice results. The dual
superconductor picture of confinement is studied by first fixing the lattice
configurations to some Abelian gauge, and then, analyzing the abelian projected configurations. 
In all the abelian gauges considered it is found that there is monopole condensation in the 
confinement phase and there is not in the de-confinement phase \cite{digiac01,digiac02,digiac03}. 
The vortex picture of confinement is studied by first fixing the gauge to Maximal Center Gauge 
and then analyzing the center projected configurations. By doing this it is observed that
these projected configurations reproduce the full string tension. Even more, this 
string tension disappears if the center vortices identified after center projection 
are removed from the lattice ensemble \cite{cendom,cendomfor}. This phenomena is called
center dominance.

The relevance of center dominance is obscured by the fact that you also obtain 
the full string tension without doing any gauge fixing \cite{ngf}. Then, center dominance seems 
of no physical relevance. Nevertheless, as it is said in \cite{ngf},
the non-triviality of center projection is related to the Maximal Center Gauge fixing
because, after doing that, the information about extended physical objects is now 
encoded in $Z_N$ local observables. So, Maximal Center Gauge is needed to identify
the vortex content of the vacuum. Doing this, Maximal Center Gauge fixing and center
projection, it is found that the obtained vortex properties extrapolate to the
continuum limit, e.g., the ratio of string tension and vortex (area) density is 
regulator independent \cite{langfeld} .

One of the drawbacks of Maximal Center Gauge is that this gauge fixing procedure suffers 
from the Gribov copies problem. This problem is associated to the structure of the 
functional to be maximized, because it has many local maximums, and then, 
the local algorithms used to find the global maximum usually ends in one of these local 
maximums (Gribov copies). As was pointed out in \cite{bornyakov,polikarpov} the Gribov copies 
problem for Maximal Center Gauge is a really severe one. Using a more
powerful algorithm (simulated annealing) to find the global maximum of the functional to be
maximized, the projected string tension is not in agreement with the physical string tension.  
This negative result has been addressed in \cite{gricop,dlcgf} arguing that
direct maximal center gauge can be understood as a best fit to a given lattice 
gauge field by a thin vortex configuration \cite{reinhardt}, being this best fit 
given by an appropriate choice of the lattice Gribov copy and, with this choice,
recovering the nice properties of the maximal center gauge fixing procedure.
Laplacian Center Gauge \cite{forcrand} was proposed as an alternative to the 
Maximal Center Gauge fixing prescription without the lattice Gribov copies 
problem. In this case, the vortex properties can be obtained with two different
procedures, looking at the points in which the gauge transformation can not be
uniquely defined or using the center projection method.

It is the purpose of this article to study how Laplacian Center Gauge identifies a thick
vortex configuration on the lattice. To this end we apply this gauge fixing procedure to 
a solution of the Yang-Mills equations of motion having vortex properties. After
gauge fixing we project to the center of the group, and then check if this solution
is seen as a thin vortex in the projected configuration. We compare the obtained
result with the one obtained using Maximal Center Gauge. Finally, we study the monopole
and vortex content of this configuration by looking at the singularities of the gauge 
fixing procedure and see how this result compare with the previous methods. 

The layout of the article is the following. In section 2 we briefly describe both 
center gauge fixing procedures, Maximal Center Gauge and Laplacian Center Gauge. 
In section 3 we show how a vortex solution appears in these gauge fixing prescriptions. 
And in section 4 we present our conclusions.

\section{Center Gauge Fixing}

In this section we briefly describe both gauge fixing procedures, Maximal Center Gauge
and Laplacian Center Gauge, for the SU(2) Yang-Mills theory.

The Maximal Center Gauge (MCG) in SU(2) lattice gauge theory is defined as the
gauge which brings link variables $U$ as close as possible to elements of its center 
$Z_2 = \pm 1$. This can be achieved by maximizing the quantity:
\bea
C=\frac{1}{VD} \sum_{n=1}^{V} \sum_{\mu =1}^{D} \frac{1}{4} 
      \mid\mbox{Tr}\;U(n,\mu)\mid^2 \hspace{0.5 cm}, \label{mesonlike}
\eea
where $V$ is the number of sites on the lattice and $D$ the number of dimensions. The 
usual procedures to maximize the functional C are local algorithms maximizing this quantity 
at each lattice point. In reference \cite{dmcgf}it is described the most used algorithm to 
perform this maximization.

The Laplacian Center Gauge (LCG) fixing prescription use the two eigenvectors with lowest
eigenvalues, $\psi^a_1(n)$ and $\psi^a_2(n)$, of the laplacian operator,
\beq
{\cal L}_{nm}^{ab}(R) = \sum_{\mu} \left( 2 \delta_{nm} \delta^{ab} - R^{ab}(n,\mu) 
                                   \delta_{m,n+\hat{\mu}}  - R^{ba}(m,\mu) 
                                   \delta_{n,m+\hat{\mu}} \right) 
\label{eq:oper}
\eeq
in presence of a gauge field $R^{ab}(n,\mu)$ in the adjoint representation of the gauge 
group, to fix completely the gauge up to the center of the SU(2) group. First, the lowest 
eigenvector, $\psi^a_1(n)$, is rotated to the ($\sigma_3$) direction in color space. This
step, Laplacian Abelian Gauge, fix the gauge up to the abelian subgroup of the SU(2) group.
This U(1) abelian freedom is fixed by imposing that the $\psi^a_2(n)$ eigenvector is further
rotated to lie in the positive ($\sigma_1,\sigma_3$) half-plane. After these two steps the
gauge is completely fixed up to the center degrees of freedom. 

\section{Gauge fixing of a Vortex Solution}

In this section we study how Laplacian Center Gauge identifies a thick vortex configuration 
on the lattice. To this end we apply this gauge fixing procedure to a solution of the Yang 
Mills classical equations of motion having the properties of a thick vortex. 
The layout of this section is the following. 
First, we review the properties of the solution we are going to work with. 
This solution was presented in \cite{vortex}. Second, we show how this solution appears after
going to Maximal Center Gauge and center projection. This result was presented in \cite{project}
and the solution appears as a thin vortex in the projected configuration. We will compare
this result with the one obtained with LCG. Third, we fix the gauge to Laplacian Center Gauge
and then we try to identify vortices in two different ways. First, by looking at the center
projected configuration, and second by looking at points in which the gauge transformation is 
not well defined. Finally, we study the monopole content of this vortex solution.

The configuration we are going to study in Maximal and Laplacian Center Gauge, is a solution
of the SU(2) Yang Mills classical equations of motion, presented in \cite{vortex}. This solution 
lives on the four dimensional torus $T^4$, with two large directions, t,x, and two small 
directions, y,z, satisfies twisted boundary conditions given by the twist vectors $\vec{k} = 
\vec{m} = (1,0,0)$, has action $S=4\pi^2$ and topological charge $|Q|=1/2$. We fix the length
of the torus in the small directions, y,z, to $l_{small}=1$. The length in the
large directions, t,x, has to be $l_{large} \gg l_{small}$ ( $l_{large}=4$ is large 
enough to obtain the desired properties of the solution). Then we have a solution
living on a four dimensional torus $T^4$ with physical sizes 
$l_{large}^2 \! \times \! l_{small}^2 \! = \! 4^2 \! \times \! 1^2$.  

The main properties of the solution are the following. By looking at the action density 
we can see that it has only one maximum and has a size approximately equal to the size  
of the torus in the small directions, y and z. The action density goes exponentially to
zero in the two large directions, t and x, while in the other two directions, y and z, never 
reaches the zero value.
This exponential fall off in t and x is the reason why $l_{large}=4l_{small}$ is big enough. 
And the most important property of this solution is that a square Wilson loop 
in the xt plane, centered at the maximum of the solution, takes the 
value $-1$ for a big enough size of the loop and is almost independent of the yz 
coordinates \cite{vortex}. Then, looking at this Wilson loop, we see a bi-dimensional object 
(because is independent of the y,z coordinates) carrying flux in an element
of the center of the group. 

To avoid any complication related to twisted boundary conditions when we gauge fix and
center project the configuration, we repeat the solution once on each direction. Then
we will have a solution in a four dimensional torus with physical size $8^2 \times 2^2$, 
with action $S= \! 2^4 \! \times \! 4 \pi^2$, topological charge 
$|Q| = \! 2^4 \! \times \! \frac{1}{2}$, and satisfying periodic boundary conditions. 
Then, we have a solution with $16$ maximums in the action density.

To obtain this solution on the lattice we use a cooling algorithm which implements
twisted boundary conditions (see \cite{minim,numer,fracton,largeN} for details on this
procedure). In this article we use three configurations obtained in lattice
sizes $N_t \! \times \! N_x \! \times \! N_y \! \times \! N_z$ with 
$N_t \! = \! N_x \! = \! 4N_s$ and $N_y \! = \! N_z \! = \! N_s \! =4,5,6$.  
As we fix the length of the torus in the small directions to be $l_{small} \! = \! 1$, 
the lattice spacing is $a \! = \! 1/N_s$. Therefore, we will be looking at the same
solution with three different resolutions, $a=0.25$, $a=0.20$ and $a=0.16$.  
Once we have these three lattice configurations, we repeat the solution in all
directions and we do not need the trick used to implement twisted boundary conditions
on the lattice. Then, we have three lattice configurations with lattice sizes 
$2N_t \! \times \! 2N_x \! \times \! 2N_y \! \times \! 2N_z$ and satisfying 
periodic boundary conditions. We label these three configurations I,II and III, for
the values of $N_s=4,5$ and $6$, respectively. From these lattice configurations the 
field strength $F_{\mu \nu}$ is obtained from the clover average of plaquettes 
$1 \! \times \! 1$ and $2 \! \times \! 2$, combined in such a way that the discretization 
errors are $O(a^4)$. And from this $F_{\mu \nu}$ we calculate all other quantities, like 
the action density or the topological charge. 

\begin{figure}
\caption{ {\footnotesize The action density S(t,x,y,z) for the III solution (lattice spacing 
$a=0.16$).  is shown as a function of x and t, and for fixed values of the y and z 
coordinates, in figure A, y and z fixed to the minimum in the action density in these
coordinates, and in figure B, y and z fixed to the maximum.} }
\vbox{ \hbox{       \vbox{ \epsfxsize=3truein \epsfysize=3.0truein \hbox{\epsffile{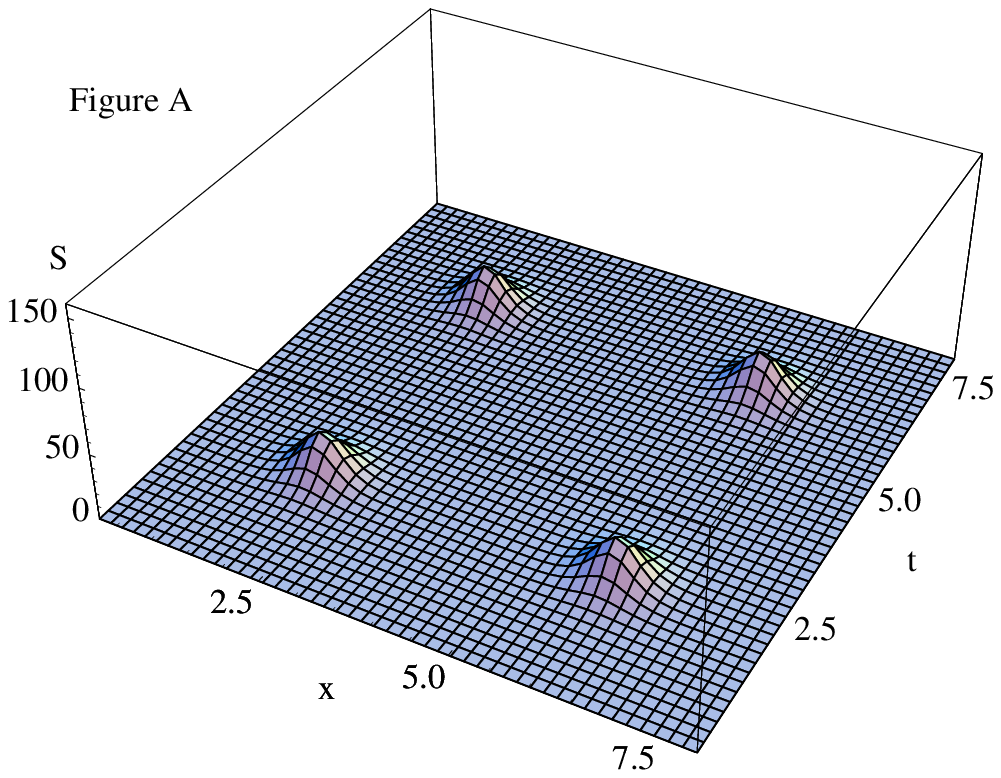} } }
             \hfill \vbox{ \epsfxsize=3truein \epsfysize=3.0truein \hbox{\epsffile{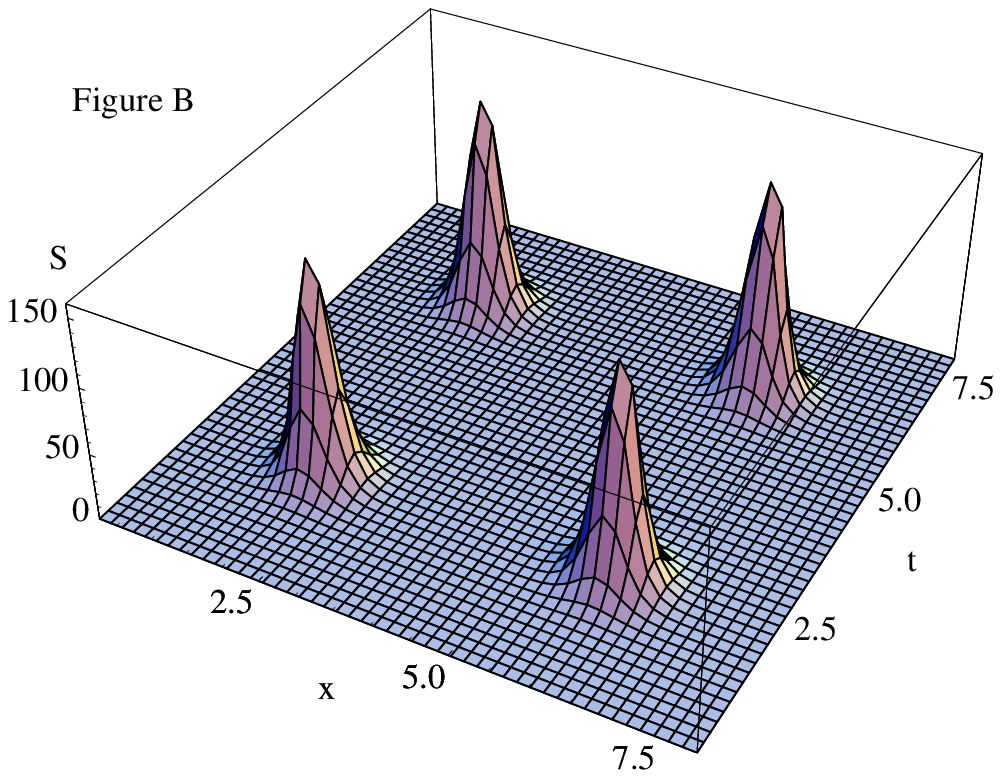} } }
 }
     }
\end{figure}

We show in figures 1A and 1B the action density for the III solution (lattice spacing $a=0.16$). 
What we plot is $S(t,x,y,z)$ for fixed values of $y$ and $z$. In figure 1B we choose 
these values to be the maximum of the action density in y,z and in figure 1A the minimum. 
We can see that the curves obtained joining the data are very smooth and also we can
figure out the dependence in y,z, for any value of y,z; we always have the picture shown
in figures 1A and 1B, but changing the height of the peak, going from the maximum value, 
shown in figure 1B, to the minimum value, shown in figure 1A. 
We also want to point out that a square Wilson loop centered in one of these 
maximums, takes the value -1 for a big enough size of the loop, in physical units
approximately equal to 2, and this value is almost independent of the $y$ and $z$ 
coordinates \cite{vortex}. 

The first thing we study is the center projected solution after going to Maximal
Center Gauge. We use the algorithm presented in \cite{dmcgf} to fix the gauge to 
Maximal Center Gauge. This is a local algorithm which maximize the functional (\ref{mesonlike}).
This procedure has the Gribov copies problem. What we made is repeat the gauge fixing
procedure several times and we take the configuration with higher value of C. In fact,
the fastest way to get the highest value of C is by going first to Laplacian Center
Gauge and then to Maximal Center Gauge. As a technical detail we say that we stop the 
gauge fixing procedure when the C quantity is stable up to the eighth significant 
digit. Once we have the gauge fixed configuration we make the center projection. 
In this case the $Z_2$ configuration is quite simple. All 
plaquettes belonging to the $xy$, $xz$, $yz$, $yt$ and $zt$ planes (planes involving
at least one of the small directions) are positives. Only in the xt plane you can find 
negative plaquettes. There are four P-vortices per xt plane, each one at the same 
location of the four maximums in the action density. All other plaquettes are positives. 
So, the vortex solution is seen, in the center projected configuration obtained after fixing
to Maximal Center Gauge, as a bi-dimensional string of negative plaquettes, this string
joining the maximums in the action density at each xt plane. 

Second, we study this solution in Laplacian Center Gauge. To fix to Laplacian Center 
Gauge we have to calculate the lowest eigenvectors of the laplacian operator. We use the 
algorithm presented in \cite{conjgrad} to obtain these vectors. We get the three 
eigenvectors with lowest eigenvalues, and we obtain that in this case the two lowest 
eigenvalues are degenerated. With these two eigenvectors we fix the gauge to Laplacian 
Center Gauge. First, we find the gauge transformation which rotates the first eigenvector 
to the third direction in color space ($\sigma_3$). 
And then, we find the abelian gauge transformation which rotates the second vector 
further to the positive $(\sigma_1,\sigma_3)$ half-plane.

We center project the LCG fixed configurations and, as before,  study the center projected
configuration. We obtain the same structure described before for the center projected 
configuration obtained after going to Maximal Center Gauge. So, laplacian center gauge and
center projection clearly identifies the vortex solution as a bi-dimensional string of
P-vortices. If we take the center projected configuration after fixing to Laplacian Abelian 
Gauge, instead of the one after LCG fixing, we do not see any structure unraveling the 
underlying vortex structure, so the fixing of the U(1) degrees
of freedom is crucial to identify the vortex properties. We also want to point out that the 
same results are obtained if we choose to fix the gauge to LCG linear combinations of the 
two lowest eigenvectors (these two are degenerated).

Monopoles and vortices can be found in Laplacian Center Gauge as defects of the gauge
fixing procedure. Then, we have to look at the first and second eigenvector and find 
the points in which you can not build the gauge transformation. In the first step, 
rotate the first eigenvector to the third direction in color space, we can find a 
singularity if $\psi^a_1(t,x,y,z)=0$. This defines lines
in four-dimensional space and these lines are identified as monopole lines. In the second
step, find the abelian gauge transformation rotating the second eigenvector 
further to the positive $(\sigma_1,\sigma_3)$ half-plane, there are singularities at points in 
which the first and second eigenvectors are parallel. This condition defines surfaces 
in four dimensional space and these surfaces are identified as vortex sheets.

\begin{figure}
\caption{ {\footnotesize The cosine of the angle between the two 
lowest eigenvectors P(t,x,y,z) for 
the III solution (lattice spacing $a=0.16$)  is shown as a function of x and t, 
and for fixed values of the y and z coordinates, in figure A, y and z fixed to 
the minimum in the action density in these coordinates, and in figure B, y and z 
fixed to the maximum.} }
\vbox{ \hbox{       \vbox{ \epsfxsize=3truein \epsfysize=3.0truein \hbox{\epsffile{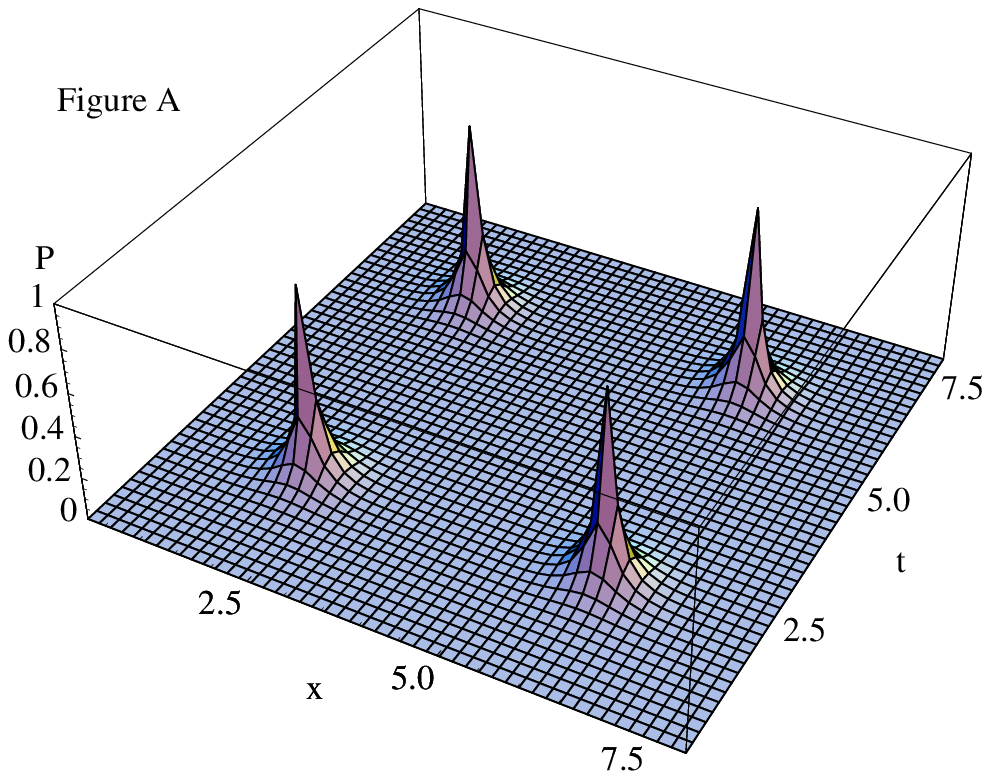} } }
             \hfill \vbox{ \epsfxsize=3truein \epsfysize=3.0truein \hbox{\epsffile{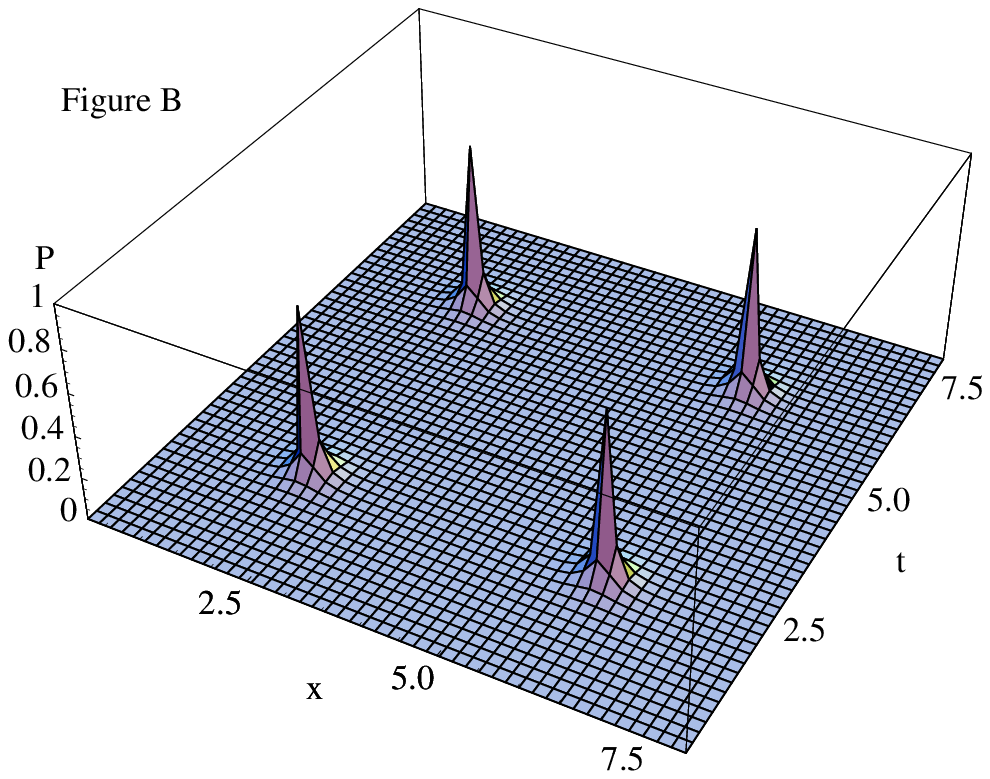} } }
 }
     }
\end{figure}

As we know that this solution is a thick vortex configuration we look at the cosine 
of the two lowest eigenvectors, P(t,x,y,z), to see how the vortex background is identified. 
In figure 2A we show P(t,x,y,z) as a function of t and x for y and z fixed to the minimum 
in the action density, and in figure 2B the same quantity but for y and z fixed to the 
maximum in the action density. The same picture is obtained for all y,z points.
We see that these two vectors are orthogonal at all points of the lattice except on  
the neighborhood of the maximum in the action density for each y,z point.
If at these points the value of P reaches the value 1 then you can not build the gauge 
transformation. This is the condition defining vortex sheets. If we join the maxima of
P(t,x,y,z) for all y,z values we obtain a surface of points with a value for the cosine
very close to 1 (always over 0.7). It seems that looking at this quantity we detect 
a vortex sheet. Note that in the case of degeneration two for the lowest eigenvector,
the vortex properties are uniquely determined because if these two vectors are parallel
at some point, any linear combination will produce parallel vectors at the same point.

Finally, we study the monopole content of this configuration. By construction, 
laplacian gauge magnetic monopoles lie on the center vortices, then, we look at
one of the $2 \! \times \! 2$ yz planes in which the vortex sheet is located 
(one of the four maximums in the action density for the x,t coordinates). 
In this case, the monopole pattern obtained depends on the choice of lowest 
eigenvector. We parameterize the possible choices with an angle, $\theta$, 
as: $\Psi_1^{'} = cos\theta \Psi_1 + sin\theta \Psi_2$,
and we look at the monopole patterns as a function of the $\theta$ angle. We always
see two monopole curves, each one going between two of the maximums in the action density  
in the y,z coordinates (positions $y_0,z_0=0.5,1.5$). For a value of $\theta$ which can 
be chosen as the origin of angles $\theta=0^o$, these two curves are straight lines in the 
y direction ($z=0.5$ and $z=1.5$ for $0 \! \le \! y \! \le \! 2$). For $\theta=45^o$ these 
curves are the two diagonals of this yz plane ($z\!=\!y$ and $z\!=\!-y+2$ for 
$0 \! \le \! y \! \le \! 2$), for $\theta=90^o$ are straight lines in the z direction 
($y=0.5$ and $y=1.5$ for $0 \! \le \! z \! \le \! 2$ ), for $\theta=135^o$ are the lines 
orthogonal to the diagonals ($z\!=\!-y\!+\!1$ and $z\!=\!y\!+\!1$ for $0 \! \le y \! \le 1$; 
$z\!=\!-y\!+\!3$ and $z\!=\!y\!-\!1$ for $1 \! \le y \! \le 2$) and for $\theta=180^o$ 
you recover the picture at $\theta=0^o$. 
So we can figure out how these curves evolve with $\theta$, the two lines in the y
direction seen at $\theta=0^o$ are deformed up to get the two diagonals at $\theta=45^o$, 
continuing then up two get the two lines in the z direction at $\theta=90^o$, then
the lines orthogonal to the diagonals at $\theta=135^o$ and finally getting the starting
pattern at $\theta=180^o$.

\section{Conclusions}

We have studied in this paper how Laplacian Center Gauge identifies a SU(2) thick vortex
configuration on the lattice. Looking at the center projected configuration obtained after 
fixing the gauge of a vortex solution to Laplacian Center gauge, we see a bi-dimensional 
string of negative plaquettes joining the maximums of the solution in the action density 
at each xt plane. This is the same result obtained by looking at the center projected 
configuration after fixing to Maximal Center Gauge. So both procedures clearly identify 
the vortex solution as a surface of P-vortices. 
We have also looked at the other way Laplacian Center Gauge can locate center vortices, 
and we have seen that looking at the possible singularities of the gauge fixing 
procedure, you obtain the same result as using Laplacian or Maximal Center Gauge  
and center projection. The candidate points to be singularities of the gauge fixing 
procedure describe the same surface detected using center projection. Nevertheless, even for this
quite simple case in which we know that there is a physical vortex, it is quite difficult
to find an interpolation procedure to state that you have an actual singularity: points
in which the lowest eigenvector of the {\bf L} operator is zero or the two lowest eigenvectors 
are parallel. This difficulty was previously pointed out in reference \cite{cvmwlgc} in which 
they use the alternative center projection procedure to locate vortices, which we have seen 
that gives the same results for the vortex solution.

It is worth to stress that, for this particular thick vortex configuration, both procedures
to locate vortices, Maximal Center projection and Laplacian Center Gauge, either with the 
collinearity condition of the two lowest eigenvectors or through the center projection
method, give the same answer. Nevertheless, this result does not allow us to conclude what
is the generic way in which Laplacian Center Gauge works, specially in realistic SU(2) 
configurations with strongly overlapping vortices. 

\section*{Acknowledgements}
I acknowledge very useful correspondence with Philippe de Forcrand.
I also acknowledge useful conversations with Margarita Garc\'{\i}a P\'erez, 
Antonio Gonz\'alez-Arroyo, John Negele, Carlos Pena, Manuel P\'erez-Victoria 
and Hugo Reinhardt. This work has been supported by the Spanish Ministerio de Educaci\'on 
y Cultura  under a postdoctoral Fellowship.


\begin{thebibliography}{10}

\bibitem{monopoles}
G. 't Hooft,
   in: {\it High Energy Physics}, Proceedings of the EPS International 
   Conference, Palermo 1975, A. Zichichi, ed., Editrice Compositori, Bologna 1976. \\
S. Mandelstan,
   Phys. Rep. 23 (1976) 245.

\bibitem{vortices}
G. 't Hooft,
   Nucl. Phys. B138 (1978) 1. \\
J. M. Cornwall,
   Nucl. Phys. B157 (1979) 392. \\
G. Mack,
   in: {\it Recent Developments in Gauge Theories}, edited by
   G. 't Hooft et al (Plenum, New York, 1980). \\
H. B. Nielsen and P. Olesen,
   Nucl. Phys. B160 (1979) 380. \\
J. Ambjorn and P. Olesen,
   Nucl. Phys. B170 (1980) 60; 265.

\bibitem{abproj}
G. 't Hooft,
   Nucl. Phys. B190 (1981) 455. 

\bibitem{digiac01}
A. Di Giacomo, B. Lucini, L. Montesi and G. Paffuti, 
Phys. Rev. D61 (2000) 034503 [hep-lat/9906024].

\bibitem{digiac02}
A. Di Giacomo, B. Lucini, L. Montesi and G. Paffuti, 
Phys. Rev. D61 (2000) 034504 [hep-lat/9906025].

\bibitem{digiac03}
J. M. Carmona, M. D'Elia, A. Di Giacomo, B. Lucini, and G. Paffuti, 
hep-lat/0103005.

\bibitem{cendom} 
L. del Debbio, M. Faber, J. Greensite and \v{S}. Olejn\'{\i}k, 
Phys. Rev. D55 (1997) 2298 [hep-lat/9610005]. 

\bibitem{cendomfor}
M. D'Elia and P. de Forcrand, 
   Phys. Rev. Lett. 82 (1999) 4582  [hep-lat/9901020].

\bibitem{ngf} 
M. Faber, J. Greensite and \v{S}. Olejn\'{\i}k, 
J. High Energy Phys. 01 (1999) 008 [hep-lat/9810008]. 

\bibitem{langfeld}
K. Langfeld, H. Reinhardt and O. Tennert
   Phys. Lett. B419 (1998) 317 [hep-lat/9710068].

\bibitem{bornyakov}
V. B. Bornyakov, D. A. Komarov, M. I. Polikarpov and A. I. Veselov,
   JETP Lett. 71 (2000) 231 [hep-lat/0002017].

\bibitem{polikarpov}
V. B. Bornyakov, D. A. Komarov and M. I. Polikarpov,
   Phys. Lett. 497 (2001) 151 [hep-lat/0009035].

\bibitem{gricop}
M. Faber, J. Greensite and \v{S}. Olejn\'{\i}k, 
   Phys. Rev. D64 (2001) 034511 [hep-lat/0103030].

\bibitem{dlcgf}
M. Faber, J. Greensite and \v{S}. Olejn\'{\i}k, 
   hep-lat/0106017.

\bibitem{reinhardt}
M. Engelhardt and H. Reinhardt,
   Nucl. Phys. B567 (2000) 249 [hep-th/9907139].

\bibitem{forcrand}
C. Alexandrou, M. D'Elia and P. de Forcrand, 
   Nucl. Phys. Proc.-Suppl 83 (2000) 4582 [hep-lat/9907028]; 
   Nucl. Phys. A663 (2000) 1031 [hep-lat/9909005].

\bibitem{dmcgf}
L. del Debbio, M. Faber, J. Giedt, J. Greensite and \v{S}. Olejn\'{\i}k,
Phys. Rev. D58 (1998) 094501 [hep-lat/9801027].

\bibitem{vansijs}
A. J. van der Sijs,
   Nucl. Phys. Proc.-Suppl 53 (1997) 535 [hep-lat/9608041]. 

\bibitem{vortex}
A. Gonz\'alez-Arroyo and \'A. Montero, 
Phys. Lett. B442 (1998) 273 [hep-th/9809037].

\bibitem{minim}
M. Garc\'{\i}a P\'erez, A. Gonz\'alez-Arroyo and B. S\"oderberg, 
Phys. Lett. B235 (1990) 117.

\bibitem{numer}
M. Garc\'{\i}a P\'erez and A. Gonz\'alez-Arroyo, 
J. Phys. A26 (1993) 2667 [hep-lat/9206016].

\bibitem{fracton}
\'A. Montero, 
J. High Energy Phys. 05 (2000) 022 [hep-lat/0004009].

\bibitem{largeN}
\'A. Montero,
Phys. Lett. B483 (2000) 309 [hep-lat/0004002].

\bibitem{project}
\'A. Montero,
Phys. Lett. B467 (1999) 106 [hep-lat/9906010].

\bibitem{conjgrad}
T. Kalkreuter and H. Simma,
Comput. Phys. Commun. 93 (1996) 33 [hep-lat/ 9507023].

\bibitem{cvmwlgc}
Ph. de Forcrand and M. Pepe,
Nucl. Phys. B598 (2001) 557 [hep-lat/0008016].

\end{thebibliography}
\end{document}